\documentclass[aps,prd,preprint,nofootinbib,superscriptaddress]{revtex4}
\usepackage{amsmath,amssymb,amsbsy,color}
\usepackage{graphicx}
\usepackage{psfrag}
\usepackage{graphicx}

\newcommand{\be}{\begin{equation}}
\newcommand{\ee}{\end{equation}}
\newcommand{\ba}{\begin{eqnarray}}
\newcommand{\ea}{\end{eqnarray}}
\newcommand{\bi}{\begin{itemize}}
\newcommand{\ei}{\end{itemize}}

\renewcommand{\>}{\rangle}  

\newcommand{\RR}{{\rm I\kern -.2em  R}}

\def\lsi{\raise0.3ex\hbox{$<$\kern-0.75em\raise-1.1ex\hbox{$\sim$}}}
\def\gsi{\raise0.3ex\hbox{$>$\kern-0.75em\raise-1.1ex\hbox{$\sim$}}}

\begin{document}

\title{
  Lattice study of meson correlators in the $\epsilon$-regime of two-flavor QCD
}

\newcommand{\NBI}{
The Niels Bohr Institute, The Niels Bohr International Academy,
Blegdamsvej 17 DK-2100 Copenhagen {\O} Denmark
}

\newcommand{\Tsukuba}{
  Graduate School of Pure and Applied Sciences, University of Tsukuba,
  Tsukuba 305-8571, Japan
}

\newcommand{\BNL}{
  Riken BNL Research Center, Brookhaven National Laboratory, Upton,
  NY11973, USA
}

\newcommand{\KEK}{
  High Energy Accelerator Research Organization (KEK),
  Tsukuba 305-0801, Japan
}

\newcommand{\GUAS}{
  School of High Energy Accelerator Science,
  The Graduate University for Advanced Studies (Sokendai),
  Tsukuba 305-0801, Japan
}

\newcommand{\NTU}{
  Physics Department, Center for Theoretical Sciences,
  and National Center for Theoretical Sciences,
  National Taiwan University, Taipei~10617, Taiwan
}

\newcommand{\YITP}{
  Yukawa Institute for Theoretical Physics,
  Kyoto University, Kyoto 606-8502, Japan
}

\author{H.~Fukaya}
\affiliation{\NBI}

\author{S.~Aoki}
\affiliation{\Tsukuba}
\affiliation{\BNL}

\author{S.~Hashimoto}
\affiliation{\KEK}
\affiliation{\GUAS}

\author{T.~Kaneko}
\affiliation{\KEK}
\affiliation{\GUAS}

\author{H.~Matsufuru}
\affiliation{\KEK}

\author{J.~Noaki}
\affiliation{\KEK}

\author{K.~Ogawa}
\affiliation{\NTU}

\author{T.~Onogi}
\affiliation{\YITP}

\author{N.~Yamada}
\affiliation{\KEK}
\affiliation{\GUAS}

\collaboration{JLQCD collaboration}
\noaffiliation


%

\begin{abstract}
  We calculate mesonic two-point functions in the $\epsilon$-regime of
  two-flavor QCD on the lattice with exact chiral symmetry.
  We use gauge configurations of size $16^3\times 32$ at $a\sim 0.11$~fm 
  generated with dynamical overlap fermions.
  The sea quark mass is fixed at around 3~MeV and the valence quark mass is
  varied in the range 1--4~MeV, both of which are in the $\epsilon$-regime.
  We find a good consistency with the expectations from the next-to-leading order
  calculation in the $\epsilon$-expansion of (partially quenched) chiral
  perturbation theory.
  From a fit we obtain the pion decay constant $F=87.3(5.6)$~MeV
  and the chiral condensate
  $\Sigma^{\overline{\mathrm{MS}}}=[239.8(4.0)\mbox{MeV}]^3$
  up to next-to-next-to-leading order contributions.
\end{abstract}

\maketitle
\section{Introduction}

In the low energy limit, the dynamics of Quantum Chromodynamics (QCD) is
dominated by the pion fields that appear as pseudo-Nambu-Goldstone bosons
associated with the spontaneous breaking of chiral symmetry.  
Although chiral Perturbation Theory (ChPT) is a powerful
effective theory in understanding their interactions, it has
many parameters, the so called 'low energy constants (LECs)', 
which should be determined either from experimental data
or preferably from direct calculations based on the underlying
theory, {\it i.e.} QCD.
The leading order of ChPT is completely controlled by the
two LECs, the chiral condensate $\Sigma$  and the pion decay
constant $F$, while at higher orders there are increasing
number of LECs.

Numerical simulation of lattice QCD offers the most
promising approach to achieve the direct calculation of LECs.
In fact, the recent advances in the simulation techniques
allow us to calculate, for instance, the pion decay constant
to a remarkable precision.
However, such calculations may contain non-trivial
systematic effects, since the infinite volume limit must be
taken before the chiral limit is approached.
The violation of chiral symmetry of lattice fermions also
seriously complicates the analysis, because ChPT itself must
be modified to incorporate the explicit breaking of chiral
symmetry \cite{Aoki:2003yv}.

Recently, an alternative approach has been proposed, that is the lattice
calculation in the $\epsilon$-regime of ChPT
\cite{Gasser:1986vb, Gasser:1987ah, Neuberger:1987zz,
Hansen:1990un, Hasenfratz:1989pk,Leutwyler:1992yt}.
In this approach, the lattice simulation is performed near the chiral limit at
a fixed volume $V$. 
Finite volume effect becomes prominent due to the long distance correlation of
the pion fields, which can be treated in a systematic way within ChPT.
Of particular interest is the region where the pion correlation length, or
the inverse pion mass $1/m_\pi$, exceeds the size of the box $L$
\begin{equation}
  \frac{1}{\Lambda_{\mathrm{QCD}}}\ll L \ll \frac{1}{m_\pi},
\end{equation}
with $\Lambda_{\rm QCD}$ the QCD scale.
In this $\epsilon$-regime, the zero-momentum mode of the pion fields has to be
treated non-perturbatively, 
and the contribution from non-zero momentum modes is expanded
in a new parameter $\epsilon$:
\begin{equation}
  \frac{m_\pi}{\Lambda_{\mathrm{cut}}} \sim 
  \frac{p^2}{\Lambda^2_{\mathrm{cut}}}\sim \epsilon^2,
\end{equation}
where $p$ denotes pion momentum and $\Lambda_{\rm cut}$ is a cutoff of ChPT.
With this expansion, the volume and topological charge dependence of the
chiral condensate, meson correlators, {\it etc.}, can be written with the same
low energy constants as in the infinite volume.

Simulating lattice QCD in the $\epsilon$-regime has several advantages over
the conventional approach.
First of all, the infinite volume limit is not necessary when approaching 
the chiral regime. 
In the p-regime, on the other hand, it is not clear
at which masses one can safely apply the chiral expansion
to extrapolate lattice data 
to the physical up and down quark masses.
This is a question that depends on the quantity of interest, and therefore
one potentially needs to work on a fairly large volume lattice to ensure the
condition $m_\pi L\gg 1$.
In the $\epsilon$-regime, pion mass is made arbitrarily small at a finite
volume. 
The $\epsilon$-expansion requires $4\pi FL\gg 1$, which does not strongly
depend on the quark mass.

One may wonder that the computational cost to simulate arbitrarily light sea
quark could be prohibitively high, but it is not the case because the
lowest-lying quark eigenvalue stays finite at the order of $1/\Sigma V$ for a
given volume $V$ (except for the exact zero-modes).
In the $\epsilon$-regime the lowest eigenvalue is even lifted by about a factor
of $N_f$, the number of flavors, according to the chiral random matrix theory.

Another major advantage of the lattice calculation in the $\epsilon$-regime is
the prominent dependence of physical observables on the topological charge and
sea quark mass. 
The best known example is the analytic predictions for the lowest-lying
eigenvalues derived from the chiral Random Matrix Theory.
Utilizing these, precise determination of the low energy constants has been
attempted through direct calculations of the low-lying Dirac eigenvalues mainly
in quenched QCD
\cite{Giusti:2002sm,Giusti:2003gf,Ogawa:2005jn}.
Another possibility is to study hadron correlators in the $\epsilon$-regime,
which has also been carried out by several groups in quenched QCD
\cite{Bietenholz:2003bj,Giusti:2003iq,Giusti:2004yp,
Fukaya:2005yg,Bietenholz:2006fj,Giusti:2007cn}.
We extend these works to unquenched QCD in this paper.
However, this is not an easy task since the exact
chiral symmetry is essential in the $\epsilon$-regime.
We therefore use the overlap-Dirac operator
\cite{Ginsparg:1981bj, Luscher:1998pq}
which realizes the exact chiral symmetry on the lattice
\cite{Ginsparg:1981bj, Luscher:1998pq}.
The overlap fermion is much more difficult to simulate
than  other lattice fermion formulations for both 
algorithmic and computational reasons.

The recent series of work by JLQCD collaboration
\cite{Fukaya:2007fb,Fukaya:2007yv, Fukaya:2007cw}
has opened a new possibility of simulating unquenched QCD in the 
$\epsilon$-regime (see also
\cite{DeGrand:2006nv,Lang:2006ab, Hasenfratz:2007yj,
DeGrand:2007tm, DeGrand:2007mi,Joergler:2007sh, 
Hasenfratz:2007qe, Jansen:2007rx} 
for other exploratory studies).
We performed two-flavor dynamical overlap fermion
simulations with the quark mass near the chiral limit,
on a $16^3\times32 $ lattice at a lattice
spacing $a\sim 0.11$~fm (determined with 
$r_0\sim 0.49$~fm \cite{Sommer:1993ce}
as an input).
We use the Iwasaki gauge action \cite{Iwasaki:1985we, Iwasaki:1984cj} with
extra Wilson fermions and ghosts 
to fix the topological charge
\cite{Izubuchi:2002pq,Vranas:2006zk,Fukaya:2006vs}. 
The sea quark mass $m$ is around 3~MeV, which is well within the
$\epsilon$-regime. 
Since the eigenvalue of the hermitian overlap-Dirac operator is bounded from
below, numerical simulation is stable even with such a small quark mass.
Comparing the Dirac spectrum with the predictions of the chiral Random Matrix
Theory, we extracted the value of the chiral condensate at the leading order
of the $\epsilon$-expansion as 
$\Sigma^{\overline{\mathrm{MS}}}=[251(7)(11)\mbox{~MeV}]^3$,
where the second error is an estimate of the systematic error due to the
next-to-leading order (NLO) effects in the $\epsilon$-expansion. 

In this paper, we use the same set of gauge configurations in the
$\epsilon$-regime to calculate the meson correlators in various channels.
The analytic predictions of ChPT for the pseudo-scalar, scalar, axial-vector,
and vector channels are known to NLO in the $\epsilon$-expansion
\cite{Damgaard:2001js,Damgaard:2002qe}, which are recently extended to the
partially quenched ChPT \cite{Damgaard:2000gh, 
Damgaard:2007ep,Bernardoni:2007hi}.  
We use these ChPT predictions to extract $\Sigma$ and $F$ at the NLO accuracy.

This paper is organized as follows.
In Section~\ref{sec:ChPT} we review the (partially quenched) ChPT predictions
for the meson correlators. 
The set-up of the numerical simulations is given in
Section~\ref{sec:simulation}. 
In Section~\ref{sec:results}, we measure the axial-vector and pseudo-scalar
correlators to extract $F$ and $\Sigma$. 
Then, some consistency checks are done using other channels and partially
quenched correlators.
Comparison of the result for $\Sigma$ is also made with that from the Dirac
eigenvalue spectrum.
Our conclusions are given in Section~\ref{sec:conclusion}.

\section{(Partially quenched) chiral perturbation theory 
in the $\epsilon$-regime at fixed topology}
\label{sec:ChPT}

In this section, we briefly review the results for the meson correlators 
calculated within (partially quenched) ChPT.
For the full details we refer the original papers
 \cite{Damgaard:2001js, Damgaard:2002qe, Damgaard:2007ep}.
Here we consider $N_v$ valence quarks with a mass $m_v$ and
$N_f=2$ degenerate sea quarks with a mass $m_s$, both in the
$\epsilon$-regime.

As a fundamental building block for the later use, 
let us define the partially quenched zero-mode partition function
\cite{Splittorff:2002eb, Fyodorov:2002wq}
at a fixed topological charge $\nu$. 
In addition to the $N_f=2$ physical quarks (of mass $m_s$), a pair of a
valence quark (of mass $m_v$) and a bosonic quark (of mass $m_b$) is
introduced for the partial quenching;
\begin{eqnarray}
\mathcal{Z}^{\rm PQ}_\nu(\mu_b|\mu_v, \mu_s)
&\equiv& 
\frac{1}{(\mu^2_s-\mu^2_v)^2}
\det \left(
\begin{array}{cccc}
K_\nu(\mu_b) & I_\nu(\mu_v) 
& I_\nu(\mu_s) & I_{\nu-1}(\mu_s)/\mu_s \\
-\mu_b K_{\nu+1}(\mu_b) & \mu_v I_{\nu+1}(\mu_v) 
& \mu_s I_{\nu+1}(\mu_s) & I_{\nu}(\mu_s) \\
\mu^2_b K_{\nu+2}(\mu_b) & \mu^2_v I_{\nu+2}(\mu_v) 
& \mu^2_s I_{\nu+2}(\mu_s) & \mu_sI_{\nu+1}(\mu_s) \\
-
\mu^3_b K_{\nu+3}(\mu_b) & \mu^3_v I_{\nu+3}(\mu_v) 
& \mu^3_s I_{\nu+3}(\mu_s) & \mu^2_sI_{\nu+2}(\mu_s) \\
\end{array}
\right),\nonumber\\
\label{eq:ZPQ}
\end{eqnarray}
where $\mu_b=m_b \Sigma V$, $\mu_v=m_v \Sigma V$, and $\mu_s=m_s \Sigma V$. 
$K_\nu$'s and $I_\nu$'s are the modified Bessel functions.
Note that, in the limit $\mu_b\to\mu_v$, (\ref{eq:ZPQ}) reduces to the
zero-mode partition function of the full $N_f=2$ theory:
\begin{eqnarray}
\lim_{\mu_b\to\mu_v}\mathcal{Z}^{\rm PQ}_\nu(\mu_b|\mu_v, \mu_s)
=
\mathcal{Z}^{\rm full}_\nu(\mu_s)\equiv \det
\left(
\begin{array}{cc}
 I_\nu(\mu_s) & I_{\nu-1}(\mu_s)/\mu_s \\
 \mu_s I_{\nu+1}(\mu_s) & I_{\nu}(\mu_s) 
\end{array}
\right).
\label{eq:Zfull}
\end{eqnarray}
Then the  partially quenched chiral condensate at finite $V$ and $\nu$ 
is given by
\begin{eqnarray}
\frac{\Sigma_{\nu}^{{\rm PQ}}(\mu_v,\mu_s)}{\Sigma}
&\equiv& 
-\lim_{\mu_b \to \mu_v}
\frac{\partial}{\partial \mu_b}
\ln \mathcal{Z}^{\rm PQ}_\nu(\mu_b|\mu_v, \mu_s)
\nonumber\\
&=&\frac{-1}{\mathcal{Z}^{\rm full}_\nu(\mu_s) (\mu^2_s-\mu^2_v)^2}\nonumber\\
&&\times \det \left(
\begin{array}{cccc}
\partial_{\mu_v} K_\nu(\mu_v) & I_\nu(\mu_v) 
& I_\nu(\mu_s) & I_{\nu-1}(\mu_s)/\mu_s \\
-\partial_{\mu_v}(\mu_v K_{\nu+1}(\mu_v)) & \mu_v I_{\nu+1}(\mu_v) 
& \mu_s I_{\nu+1}(\mu_s) & I_{\nu}(\mu_s) \\
\partial_{\mu_v}(\mu^2_v K_{\nu+2}(\mu_v)) & \mu^2_v I_{\nu+2}(\mu_v) 
& \mu^2_s I_{\nu+2}(\mu_s) & \mu_sI_{\nu+1}(\mu_s) \\
-\partial_{\mu_v}(\mu^3_v K_{\nu+3}(\mu_v)) & \mu^3_v I_{\nu+3}(\mu_v) 
& \mu^3_s I_{\nu+3}(\mu_s) & \mu^2_sI_{\nu+2}(\mu_s) \\
\end{array}
\right).
\end{eqnarray}
It is not difficult to see that in the $\mu_v \to \mu_s$ limit, 
the partially quenched 
condensate reduces to the one in the full theory,
\begin{eqnarray}
\frac{\Sigma_{\nu}^{{\rm PQ}}(\mu_s,\mu_s)}{\Sigma}
&=&  \frac{\Sigma_{\nu}^{{\rm full}}(\mu_s)}{\Sigma}
\equiv 
\frac{1}{2}\frac{\partial}{\partial \mu_s}\ln 
\mathcal{Z}^{\rm full}_\nu(\mu_s).
\end{eqnarray}
In the following, we will also use a second-derivative
\begin{eqnarray}
\frac{\Delta \Sigma_{\nu}^{{\rm PQ}}(\mu_v,\mu_s)}{\Sigma}
&\equiv& 
\frac{\lim_{\mu_b \to \mu_v}
\partial_{\mu_b} \partial_{\mu_v} 
\mathcal{Z}^{\rm PQ}_\nu(\mu_b|\mu_v, \mu_s)}
{\mathcal{Z}^{\rm full}_\nu(\mu_s)}.
\end{eqnarray}

First we present the two-point correlation functions of the 
flavored pseudo-scalar and scalar operators,
$P^a(x)=\bar{q}(x)\tau^a \gamma_5 q(x)$ and
$S^a(x)=\bar{q}(x)\tau^a q(x)$,
where $\tau^a$ denotes a generator of $SU(N_v)$ group that the valence quark
field $q(x)$ belongs to.

The expressions for the correlators
$\<P^a(x)P^a(0)\>$ and $\<S^a(x)S^a(0)\>$
in the partially quenched ChPT is known to ${\cal O}(\epsilon^2)$ 
(no sum over $a$) \cite{Damgaard:2007ep}:
\begin{eqnarray}
\label{eq:Pmult}
C_P(t)&\equiv&
\int d^3 x \langle P^a(x)P^a(0)\rangle 
=\frac{1}{2}
\frac{L^3\Sigma_{\rm eff}^2}{\mu_v^{\rm eff}}
\frac{\Sigma_{\nu}^{{\rm PQ}}(\mu_v^{\rm eff}, \mu_s^{\rm eff})}{\Sigma}
-\frac{1}{2}\left[
\frac{2\Sigma^2}{F^2}
\frac{\Delta\Sigma_{\nu}^{{\rm PQ}}(\mu_v,\mu_s)}{\Sigma}
\right.
\nonumber\\
&&
\left.
+\frac{\Sigma^2}{F^2}\frac{\partial_{\mu_v}
\Sigma_{\nu}^{{\rm PQ}}(\mu_v, \mu_s)}{\Sigma}
-\frac{\Sigma^2}{F^2}
\frac{4}{\mu_v^2-\mu_s^2}\left(
\frac{\mu_v\Sigma_{\nu}^{{\rm PQ}}(\mu_v, \mu_s)}{\Sigma}
-\frac{\mu_s\Sigma_{\nu}^{{\rm full}}(\mu_s)}{\Sigma}
\right)
\right]Th_1(t/T),\nonumber\\
\\
\label{eq:Smult}
C_S(t)&\equiv&\int d^3 x \langle S^a(x)S^a(0)\rangle 
=
\frac{L^3\Sigma_{\rm eff}^2}{2}
\frac{\partial_{\mu_v}
\Sigma_{\nu}^{{\rm PQ}}(\mu_v^{\rm eff}, \mu_s^{\rm eff})}{\Sigma}
-\frac{1}{2}\left[
\frac{2\Sigma^2}{F^2}
\frac{\nu^2}{\mu_v^2}
\right.
\nonumber\\
&&
\left.
+\frac{\Sigma^2}{F^2}\frac{1}{\mu_v}\frac{
\Sigma_{\nu}^{{\rm PQ}}(\mu_v, \mu_s)}{\Sigma}
-\frac{\Sigma^2}{F^2}
\frac{4}{\mu_v^2-\mu_s^2}\left(
\frac{\mu_v\Sigma_{\nu}^{{\rm PQ}}(\mu_v, \mu_s)}{\Sigma}
-\frac{\mu_s\Sigma_{\nu}^{{\rm full}}(\mu_s)}{\Sigma}
\right)
\right]Th_1(t/T).\nonumber\\
\end{eqnarray}
The operators are separated in time by $t$ and projected onto zero spatial
momentum.
At the NLO an ``effective'' chiral condensate
$\Sigma_{\mathrm{eff}}$, 
\begin{eqnarray}
\label{eq:Sigmaeff}
\Sigma_{\rm eff}=\Sigma 
\left(1+\frac{3\beta_1}{2F^2V^{1/2}}\right),
\end{eqnarray}
appears with a combination $\mu_i^{\rm eff}= m_i\Sigma_{\mathrm{eff}}V$.
The parameter $\beta_1$ is the so-called shape coefficient that depends on the
shape of the box. In our numerical study, $\beta_1=0.0836$.
In Eqs. (\ref{eq:Pmult}) and (\ref{eq:Smult}),
the time dependence is simply written
by a quadratic function $h_1(t/T)$ as
\begin{eqnarray}
h_1(t/T) &\equiv& 
\frac{1}{2}
\left[\left(\frac{t}{T}-\frac{1}{2}\right)^2-\frac{1}{12}\right].
\end{eqnarray}

Next, we consider the flavored axial-vector and vector operators 
$A^a_0(x)=\bar{q}(x)\tau^a \gamma_0 \gamma_5 q(x)$ and
$V^a_0(x)=\bar{q}(x)\tau^a \gamma_0 q(x)$.
These correlators to ${\cal O}(\epsilon^2)$ in the $N_f=2$ unquenched ChPT are
\cite{Damgaard:2002qe}
\begin{eqnarray}
  C_A(t)\equiv\int d^3x 
  \langle A^a_0(x) A^a_0(0) \rangle 
  & = & -\frac{F^2}{2 T}\left\{{\cal J}^0_+ 
    +\frac{2}{F^2} \left(\frac{\beta_1}{V^{1/2}} {\cal J}^0_+ 
      -\frac{T^2}{V} k_{00} {\cal J}^0_- \right) \right.\nonumber\\
  & & \hspace*{1.75cm}\left. +\frac{4 \mu_s}{F^2} 
    \frac{\Sigma_{\nu}^{{\rm full}}(\mu_s)}{\Sigma} \frac{T^2}{V}h_1(t/T)
  \right\},
  \label{eq:fullaxial}
 \\
 C_V(t)\equiv\int d^3 x 
 \langle V^a_0(x) V^a_0(0) \rangle
 & = & -\frac{F^2}{2 T}\left\{{\cal J}^0_- 
   +\frac{2}{F^2} \left(\frac{\beta_1}{V^{1/2}} {\cal J}^0_- 
     -\frac{T^2}{V} k_{00} {\cal J}^0_+ \right) \right\},
 \label{eq:fullvector}
\end{eqnarray}
where $k_{00}$ is another numerical factor depending on the shape 
of the box. In our numerical study, $k_{00}=0.08331$.
${\cal J}^0_\pm$'s are defined by
\begin{eqnarray}
 {\cal J}^0_{+} & \equiv & 
 \frac{1}{3} \left( 2 + 2 \left[ \frac{\partial_{\mu_s}
\Sigma^{\rm full}_\nu(\mu_s^{\rm eff})}{\Sigma} 
 + 2 \left(\frac{\Sigma^{\rm full}_\nu(\mu^{\rm eff}_s)}{\Sigma}\right)^2 
 + \frac{1}{\mu^{\rm eff}_s} \frac{\Sigma^{\rm full}
_\nu(\mu^{\rm eff}_s)}{\Sigma} 
 - 2 \frac{\nu^2}{(\mu^{\rm eff}_s)^2} \right]
 \right), 
 \label{J0p}
 \\
 {\cal J}^0_{-} & \equiv & 
\frac{1}{3} \left( 4
- 2\left[ \frac{\partial_{\mu_s}
\Sigma^{\rm full}_\nu(\mu_s^{\rm eff})}{\Sigma} 
 + 2 \left(\frac{\Sigma^{\rm full}_\nu(\mu^{\rm eff}_s)}{\Sigma}\right)^2 
 + \frac{1}{\mu^{\rm eff}_s} \frac{\Sigma^{\rm full}
_\nu(\mu^{\rm eff}_s)}{\Sigma} 
 - 2 \frac{\nu^2}{(\mu^{\rm eff}_s)^2} \right]
\right). 
\label{J0m}
\end{eqnarray}

It should be noted that the axial-vector and vector correlators
are sensitive to $F$, as it appears as an overall constant.
Their time dependence and other factors represent the NLO effect.
For the pseudo-scalar and scalar correlators, on the other hand, $\Sigma$ 
determines the overall constant and $F$ appears only in the NLO correction
term.
Therefore, by using both types of correlators the two LECs can be extracted
with a good sensitivity.

\section{Lattice simulations}
\label{sec:simulation}

We summarize the setup of our numerical simulations.
Details of the configuration generation and the eigenvalue
calculations are given in \cite{Fukaya:2007yv}.


Our lattice is $16^3\times 32$ at a lattice spacing
$a=$ 0.1111(24)~fm determined from 
the Sommer scale $r_0=0.49$~fm as an input \cite{Sommer:1993ce}.
We employ the overlap fermion \cite{Neuberger:1997fp,Neuberger:1998wv},
defined by the Dirac operator
\begin{equation}
\label{eq:ov}
D(m)=\left(m_0 + \frac{m}{2}\right)
+\left(m_0 - \frac{m}{2}\right)\gamma_5 \mbox{sgn}
[H_W(-m_0)]
\end{equation}
for a quark mass $m$.
$H_W(-m_0)\equiv\gamma_5D_W(-m_0)$ denotes the standard Hermitian
Wilson-Dirac operator at a large negative mass $-m_0$.
We choose $m_0=1.6$ throughout this work. 
We remark that the mass parameters are given in the
lattice unit unless otherwise stated.
We employ the Iwasaki action 
\cite{Iwasaki:1985we,Iwasaki:1984cj}
for the gauge field at
$\beta=2.35$ with an additional determinant factor
$\det[H_W^2/ (H_W^2+m_t^2)]$ in the partition function produced by extra
Wilson fermions and twisted-mass ghosts 
\cite{Izubuchi:2002pq,Vranas:2006zk,Fukaya:2006vs}
(we set $m_t=0.2$).
With this choice the global topological charge does not change its value
during the molecular dynamics updates of the Hybrid Monte Carlo algorithm.
Fixing topology in this way is desirable for the study of the
$\epsilon$-regime of QCD, since the analytical expressions of ChPT are given
at a fixed topological sector.
In this work, we take the trivial topological sector $\nu=0$ only.

The sign function in (\ref{eq:ov}) is approximated by a rational function with
Zolotarev's coefficients after projecting out a few lowest eigenmode's
contribution. 
With 10 poles the accuracy of the sign function is $10^{-(7-8)}$.
Therefore, the violation of the chiral symmetry due to the lattice action is
negligible in our work.


In this work, we use the gauge configurations generated at sea quark mass
$m=0.002$, which corresponds to $\sim 3$~MeV.
For this value the parameter $m\Sigma V$ is about 0.556, and the system is well
within the $\epsilon$-regime.
As we decrease the quark mass from the $p$-regime ($m\Sigma V\gg 1$) to the
$\epsilon$-regime ($m\Sigma V\lesssim 1$), the lowest eigenvalue of 
$D(m)^\dagger D(m)$, $\simeq\lambda_1^2+m^2$, is bounded from below by a
repulsion of the lowest eigenvalue from zero due to the fermion determinant
$\prod_k(\lambda_k^2+m^2)$.
The condition number of the operator $D(m)^\dagger D(m)$, and thus the
computational cost for its inversion, saturates near the boarder between the
$p$-regime and the $\epsilon$-regime.
As a consequence, for example, the number of multiplication of the
Wilson-Dirac operator $D_W$ needed per trajectory at $m=0.002$ is only about
1.5 times greater than that at 10 times heavier sea quark mass 
\cite{Fukaya:2007yv}.
We have accumulated 4,600 HMC trajectories after discarding 400 trajectories
for thermalization. 
The numerical cost is about one hour per trajectory on a half rack 
(512 nodes) of the IBM BlueGene/L (2.8~TFlops peak performance).

At every 10 trajectories, we calculate the meson correlators at four values of
valence quark masses $m=0.0005$, 0.0010, 0.0020, and 0.0030 ranging 
1--4~MeV in the physical unit.
For the inversion of the overlap-Dirac operator we use the multi-shift
Conjugate Gradient (CG) solver to calculate all the valence quark propagators
simultaneously. 
The solver is accelerated by projecting out the subspace spanned by 50 pairs of
lowest-lying eigenmodes.
(The eigenvalues of $D(0)$ form a pair with their complex
conjugate; the eigenvector of the counterpart is produced by multiplying
$\gamma_5$.) 
With this projection, the solver performance is an order of magnitude better
and roughly independent of the quark mass.
These eigenmodes of $D(0)$ has been
calculated using the implicitly restarted Lanczos algorithm
and stored on disks for studying the eigenvalue distribution
\cite{Fukaya:2007fb,Fukaya:2007yv}. 

We compute the meson correlators
with the Low Mode Averaging
(LMA) technique \cite{DeGrand:2004qw,Giusti:2004yp}.
Using the $N_{ep}$ pairs of lowest-lying eigenmodes, 
we decompose the quark propagator $D(m)^{-1}$
into the low-mode contribution $[D(m)]_{low}^{-1}$ and the rest
$[D(m)]_{high}^{-1}$ as
\begin{eqnarray}
  \label{eq:decompose}
  D(m)^{-1}(x,y) & = & [D(m)]_{low}^{-1}(x,y) + [D(m)]_{high}^{-1}(x,y) 
  \nonumber\\
  & = &
  \sum_{k=1}^{N_{ep}} \left[
    \frac{u_k(x)u_k^\dagger(0)}{(1-m/2m_0)\lambda_k+m} +
    \frac{\gamma_5u_k(x)u_k^\dagger(0)\gamma_5}{(1-m/2m_0)\lambda_k^\ast+m}
  \right]
  + [D(m)]_{high}^{-1}(x,y),\nonumber\\
\end{eqnarray}
where $u_k(x)$ is the eigenvector of $D(0)$ 
associated with its eigenvalue $\lambda_k$.
While the high-mode contribution have to be obtained by using the CG solver
for a fixed source point $y$, the low-mode contribution can be calculated from
the 50+50 low-lying modes for any source and sink points without extra cost.
With the LMA technique, we average over the source point for a part of 
the meson correlator that is purely composed of $[D(m)]_{low}^{-1}$.
For other contributions, we simply use a fixed source $y$ at the origin.

\begin{figure*}[tbp]
  \centering
  \includegraphics[width=8cm]{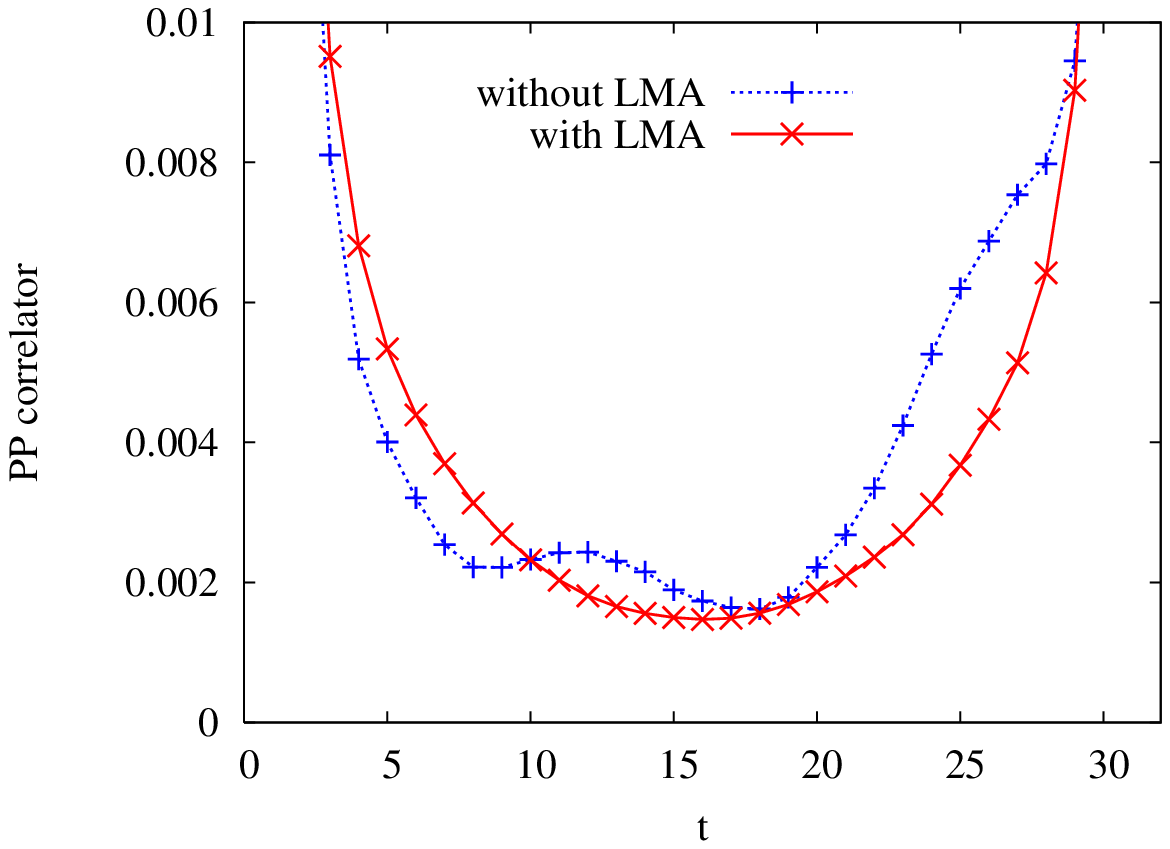}
  \includegraphics[width=8cm]{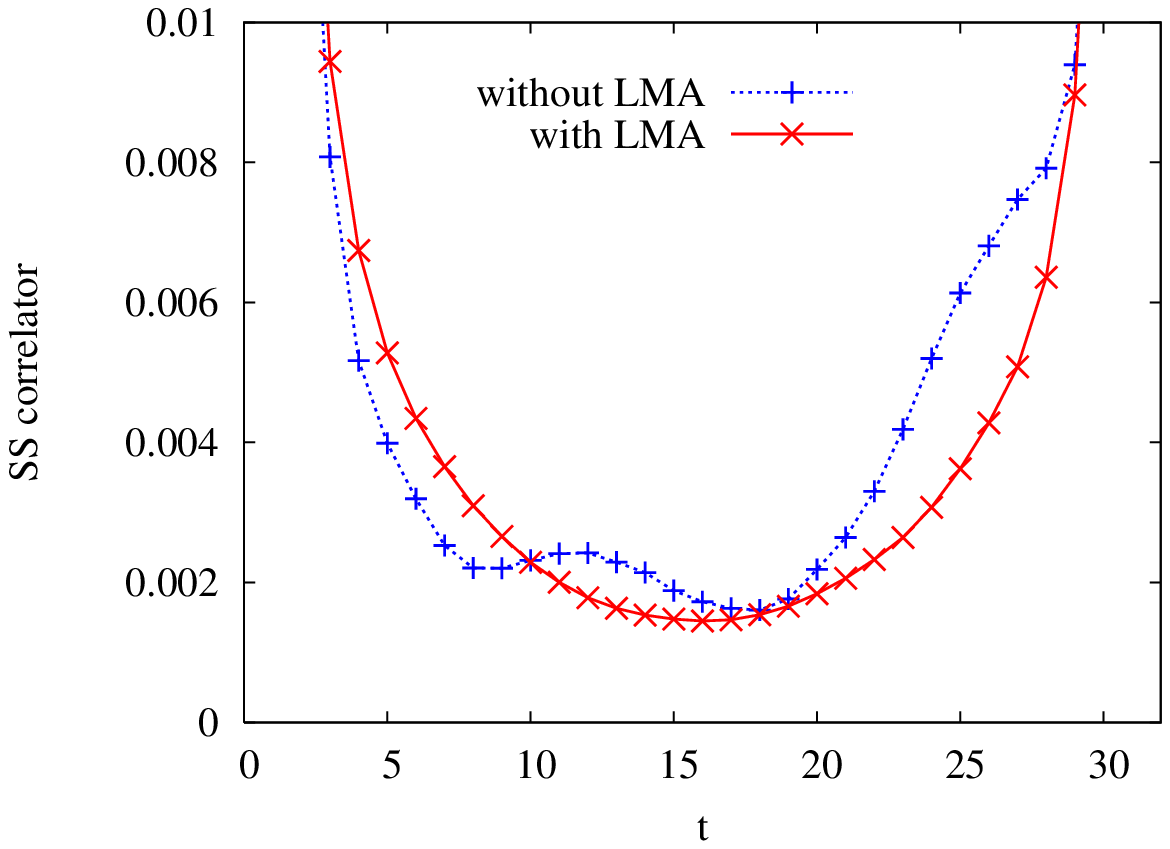}
  \caption{
    Pseudo-scalar (left) and scalar (right) correlators with (solid curve) and
    without (dotted curve) the low-mode averaging.
    Correlators are calculated on a single gauge configuration at $m=0.002$.
  }
  \label{fig:LMA-P-S}
\end{figure*}

\begin{figure*}[tbp]
  \centering
  \includegraphics[width=8cm]{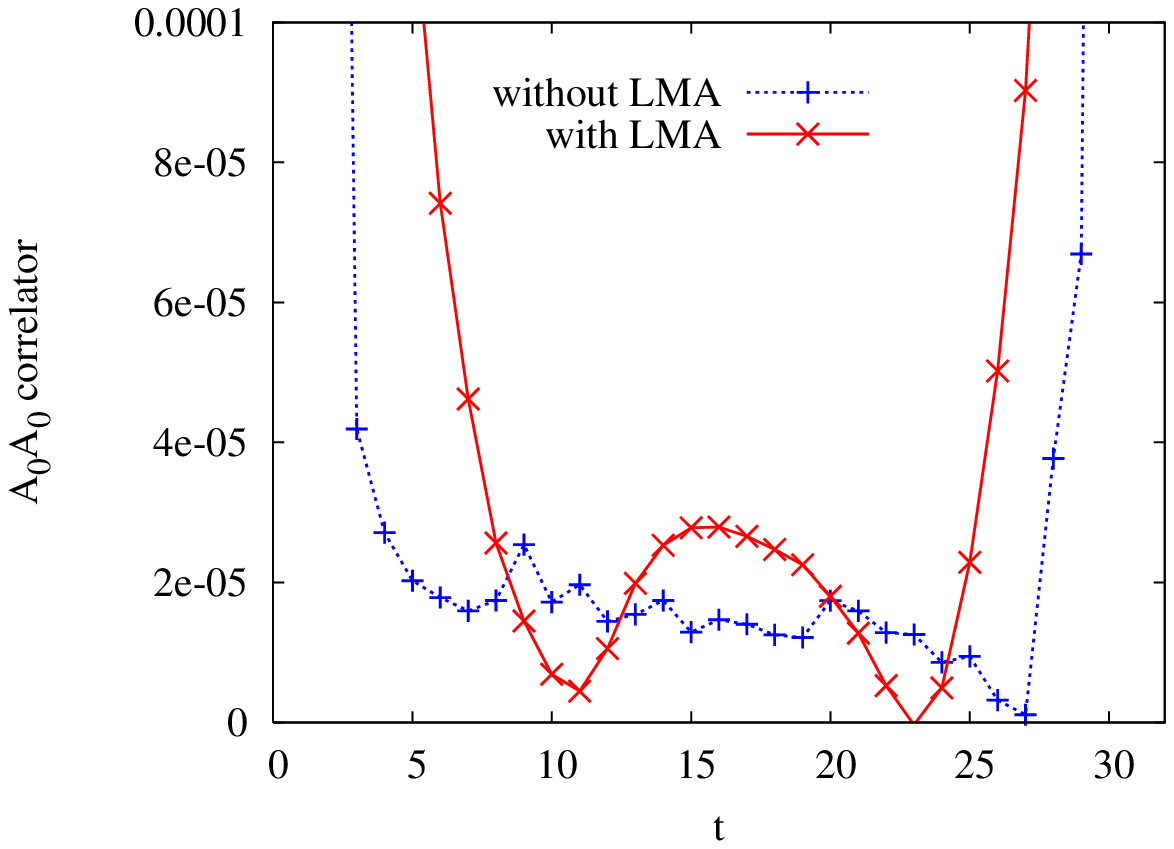}
  \includegraphics[width=8cm]{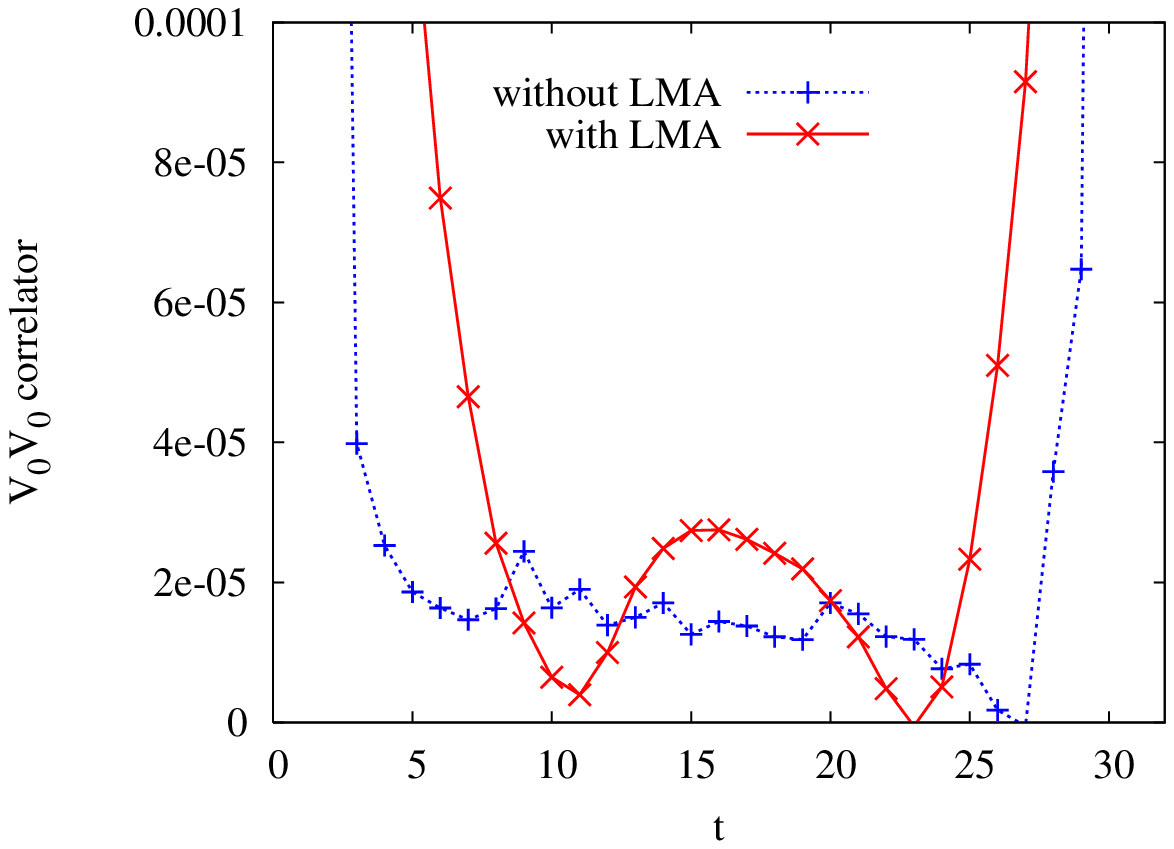}
  \caption{
    Axial-vector (left) and vector (right) correlators with (solid curve) and
    without (dotted curve) the low-mode averaging.
    Correlators are calculated on a single gauge configuration at $m=0.002$.
  }
  \label{fig:LMA-A-V}
\end{figure*}

The LMA technique is effective to improve the statistical signal when the
correlator of interest is dominated by the low-mode contribution.
In Figure~\ref{fig:LMA-P-S} we demonstrate the improvement by taking
pseudo-scalar and scalar correlators calculated on a single gauge
configuration as an example.
Some wiggle observed without LMA is completely washed out with LMA, and smooth
curve is obtained.
A similar comparison is shown in Figure~\ref{fig:LMA-A-V} for the axial-vector
and vector correlators, for which the improvement with LMA is marginal.
This indicates that these correlators are not simply dominated by the
low-lying modes.
Note that the magnitude of these correlators is two orders of magnitude
smaller than pseudo-scalar and scalar correlators.

\begin{figure}[tbp]
  \centering
  \includegraphics[width=12cm]{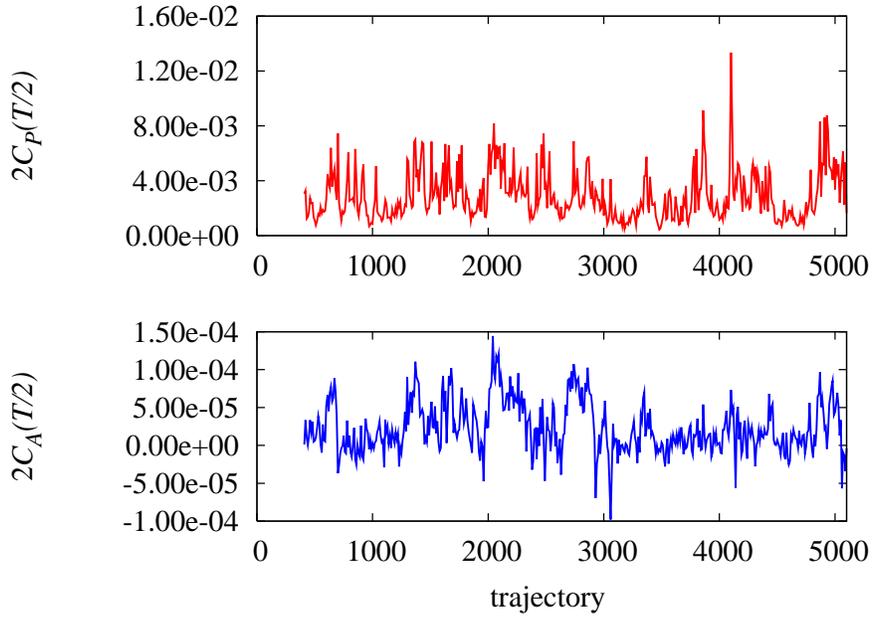}
  \caption{
    The Monte Carlo history of the pseudo scalar correlator (top)
    and the axial vector correlator (bottom) at $t=T/2=16$.
  }
  \label{fig:auto}
\end{figure}

For the statistical analysis, we use the jackknife method with a bin size 20,
which corresponds to 200 HMC trajectories.
With this choice the statistical error saturates for the calculation of the
averaged low-lying eigenvalues $\langle\lambda_k\rangle$ as studied in
\cite{Fukaya:2007yv}. 
Since the low-lying modes reflect the long-distance physics (zero-modes of
pion fields in the language of ChPT), these quantities are expected to have
the longest auto-correlation time among other physical quantities.
In Figure~\ref{fig:auto}, we plot the Monte Carlo history of the quantities of
interest, {\it i.e.} the pseudo-scalar and axial-vector correlators at the
largest time separation $t=T/2=16$, $C_P(T/2)$ and $C_A(T/2)$ 
(see the definition given later).
We observe that the 200-trajectory gives a reasonable range of the
auto-correlation.
However, the integrated auto-correlation time calculated following the
definition in \cite{Luscher:2005rx} is substantially shorter:
$37(12)$ and $47(16)$ HMC trajectories for $C_P(T/2)$ and $C_A(T/2)$,
respectively. 
We therefore conclude that the binsize of 20 is a conservative choice.

\section{Numerical results} 
\label{sec:results} 

In this section we explain the fit of our data to the (partially quenched)
ChPT formulae in Eqs.(\ref{eq:Pmult}), (\ref{eq:Smult}), 
(\ref{eq:fullaxial}), and (\ref{eq:fullvector}). 

First, we study the axial-vector current correlator at
$m_v=m_s=0.002$ (full QCD point).
Since we use the local axial current $A^a_0(x)$,
which is not a conserved current on the lattice, 
we need a finite renormalization to relate the lattice current
to the continuum current $\mathcal{A}^a_0(x)$ 
as $\mathcal{A}^a_0(x)=Z_A A^a_0(x)$. 
We calculated the renormalization factor $Z_A$ 
non-perturbatively through the axial Ward Identity and 
obtain $Z_A=1.3513(13)$.
In the following, the values quoted for $F$ include this $Z$-factor,
neglecting its tiny statistical error.

\begin{figure}[tbp]
  \centering
  \includegraphics[width=12cm]{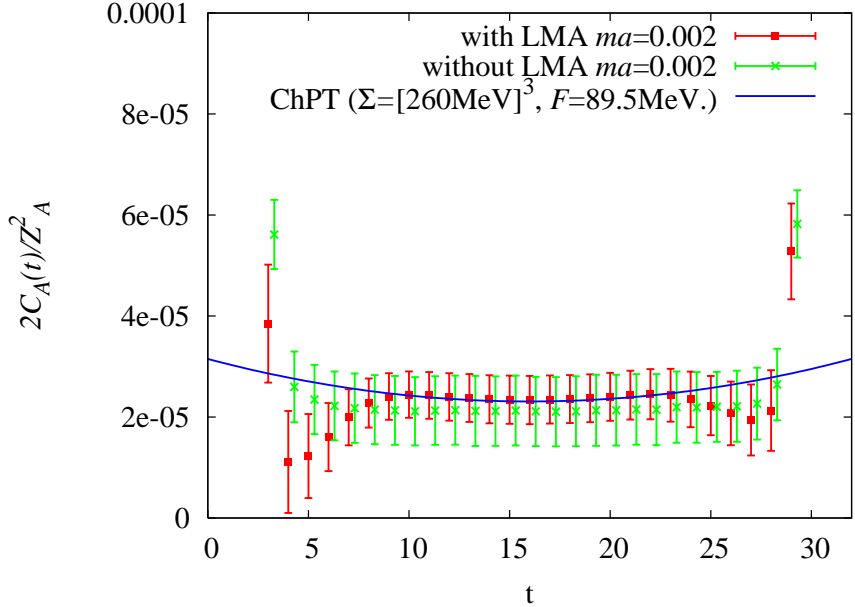}
  \caption{
    Axial-vector correlator in the $\epsilon$-regime.
    Filled square symbols denote the low-mode averaged correlator
    while the crosses are not averaged.
    The solid curve shows the best fit with the ChPT formula.
  }
  \label{fig:axial}
\end{figure}

Figure~\ref{fig:axial} shows the axial-vector current correlator with and
without the LMA technique.
Although the improvement by LMA is marginal for this channel, the statistical
error is reduced by about 30\%.
We fit the data to the ChPT formula (\ref{eq:fullaxial}) taking $F$ and
$\Sigma$ as free parameters.
With a fitting range $t\in [12, 20]$ we obtain the solid curve shown in 
Figure~\ref{fig:axial}.
Roughly speaking, the overall magnitude (the constant piece) determines $F$,
while the curvature (or the term proportional to $h_1(t/T)$) gives $\Sigma$.
We obtain $\Sigma \sim [260(32)\mathrm{~MeV}]^3$ and $F\sim 90(6)$~MeV.
The statistical error is large for $\Sigma$ (about 30\%), because it is
extracted from a tiny curvature.

\begin{figure}[tbp]
  \centering
  \includegraphics[width=12cm]{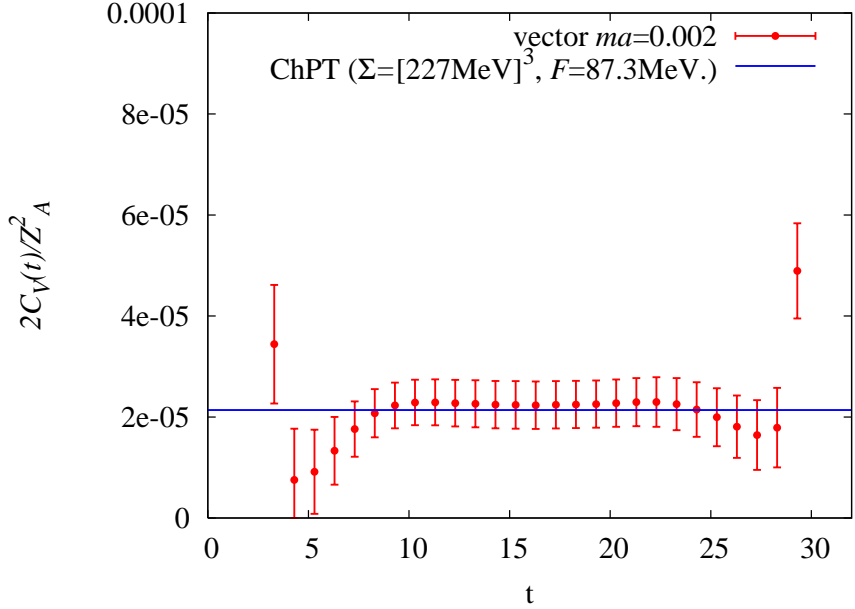}
  \caption{
    Vector correlator in the $\epsilon$-regime.
    The solid curve is the ChPT prediction
    with inputs $\Sigma = [227.6(3.7)\mathrm{~MeV}]^3$ and
    $F = 87.3(5.6)\mathrm{~MeV}$ (No free parameter left).
  }
  \label{fig:vector}
\end{figure}

A similar plot is obtained for the vector channel as shown in
Figure~\ref{fig:vector}. 
At the leading order, {\it i.e.} in the massless limit at finite $V$,
axial-vector and vector correlators become identical because of the exact
chiral symmetry; the difference arises due to the
zero-mode integrals when $\mu$ is finite.
The line in the plot shows the formula (\ref{eq:fullvector}) with the
parameters $F$ and $\Sigma$ obtained from the fit of the axial-vector and
pseudo-scalar channels (see Section~\ref{sec:LEC}).
It shows a remarkable consistency.

\begin{figure}[tbp]
  \centering
  \includegraphics[width=12cm]{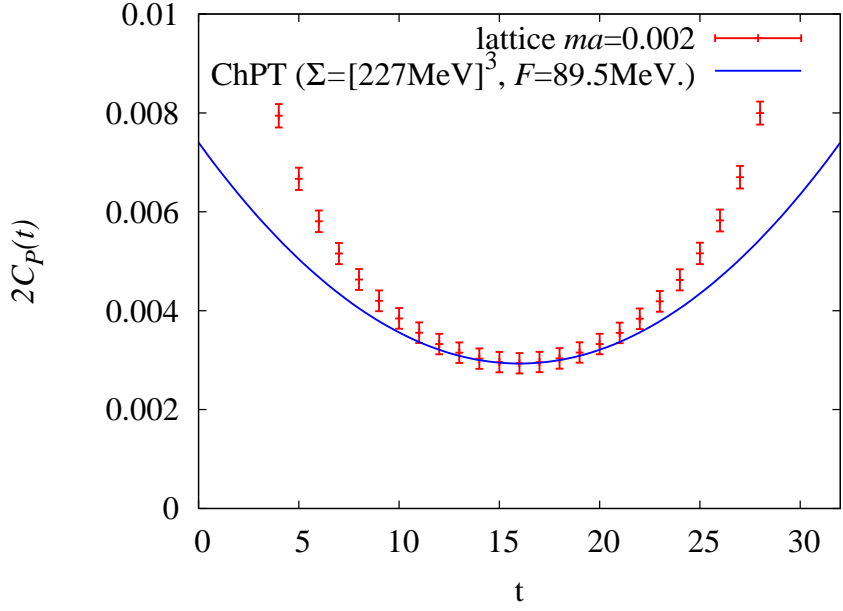}
  \caption{
    Pseudo-scalar correlator in the $\epsilon$-regime.
    The solid curve represents a fit with the NLO ChPT formula.
  }
  \label{fig:pion}
\end{figure}

Next, we consider the pseudo-scalar channel at $m_v=m_s=0.002$.
The lattice data obtained with the LMA technique are shown in
Figure~\ref{fig:pion}. 
With $F$ obtained via the axial vector correlator as an input,
we compare the pseudoscalar channel with the chiral perturbation theory
formula at the NLO (\ref{eq:Pmult}).
The motivation for this choice is that the pseudo-scalar channel itself does
not have a good sensitivity on $F$, as it appears only at the NLO.
From a fit in the range $t\in [12,20]$ we obtain the chiral condensate as
$\Sigma \sim [227(4)\mathrm{~MeV}]^3$.

\begin{figure}[tbp]
  \centering
  \includegraphics[width=12cm]{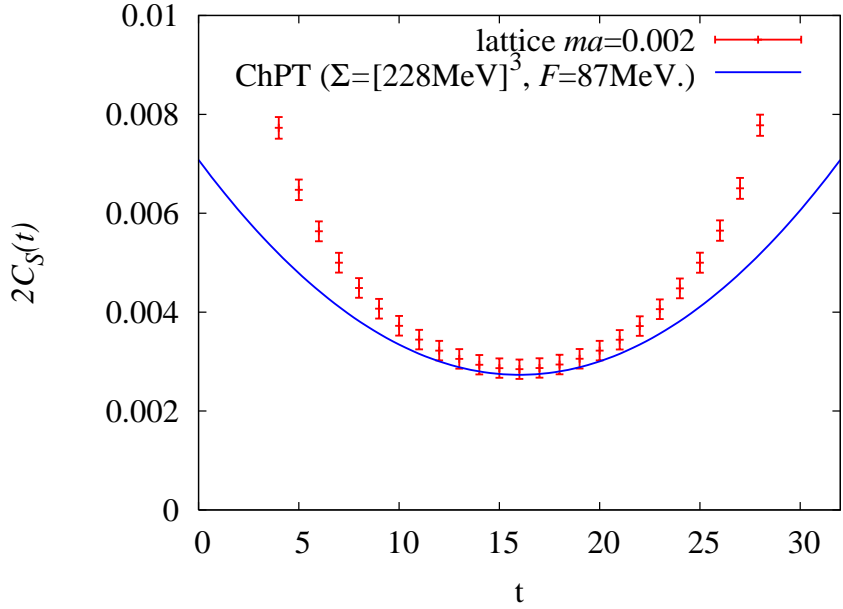}
  \caption{
    Scalar correlator in the $\epsilon$-regime.
    The solid curve represents a fit with the NLO ChPT formula.
  }
  \label{fig:scalar}
\end{figure}

A similar result can be obtained for the scalar channel as plotted in
Figure~\ref{fig:scalar}. 
We draw a curve representing the ChPT formula (\ref{eq:Smult}) with the
parameters obtained from the fit of the axial-vector and
pseudo-scalar channels (see Section~\ref{sec:LEC}).
Again for this channel, the data are consistent with the expectation within
the statistical error.

\begin{figure}[tbp]
  \centering
  \includegraphics[width=12cm]{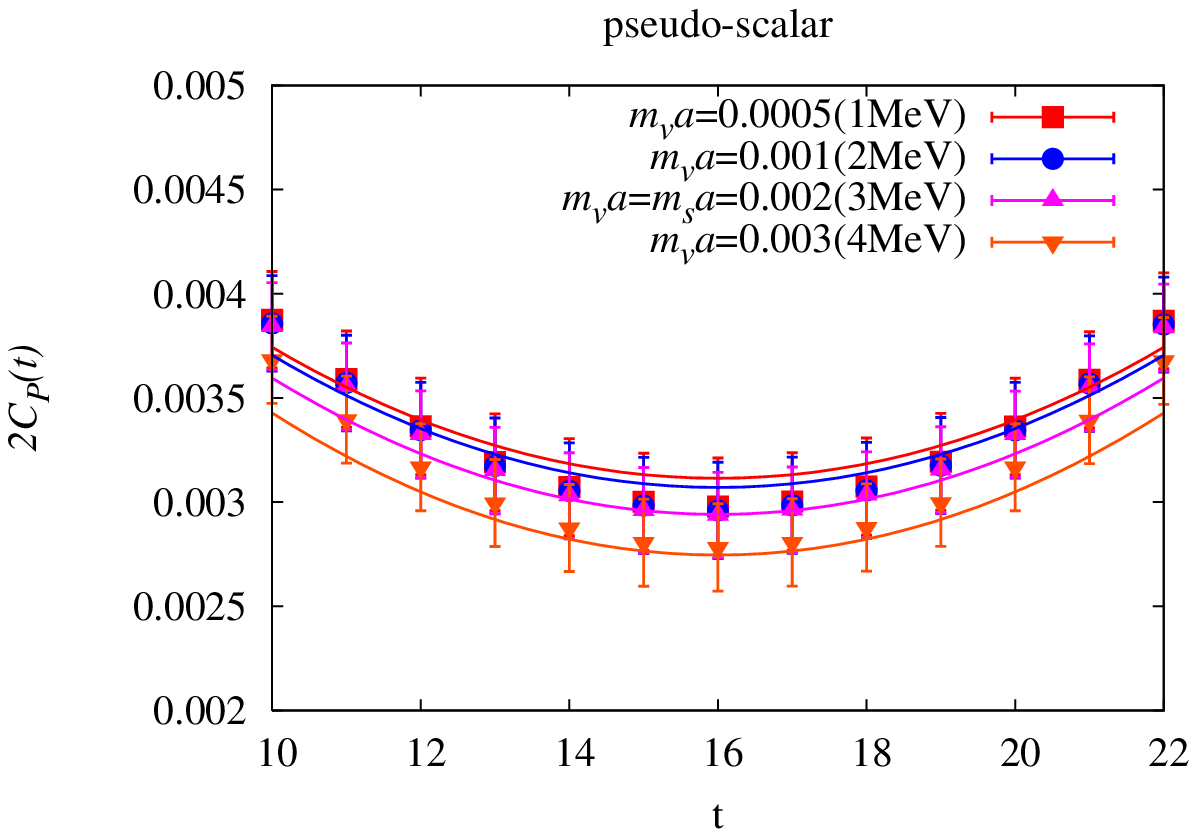}
 \includegraphics[width=12cm]{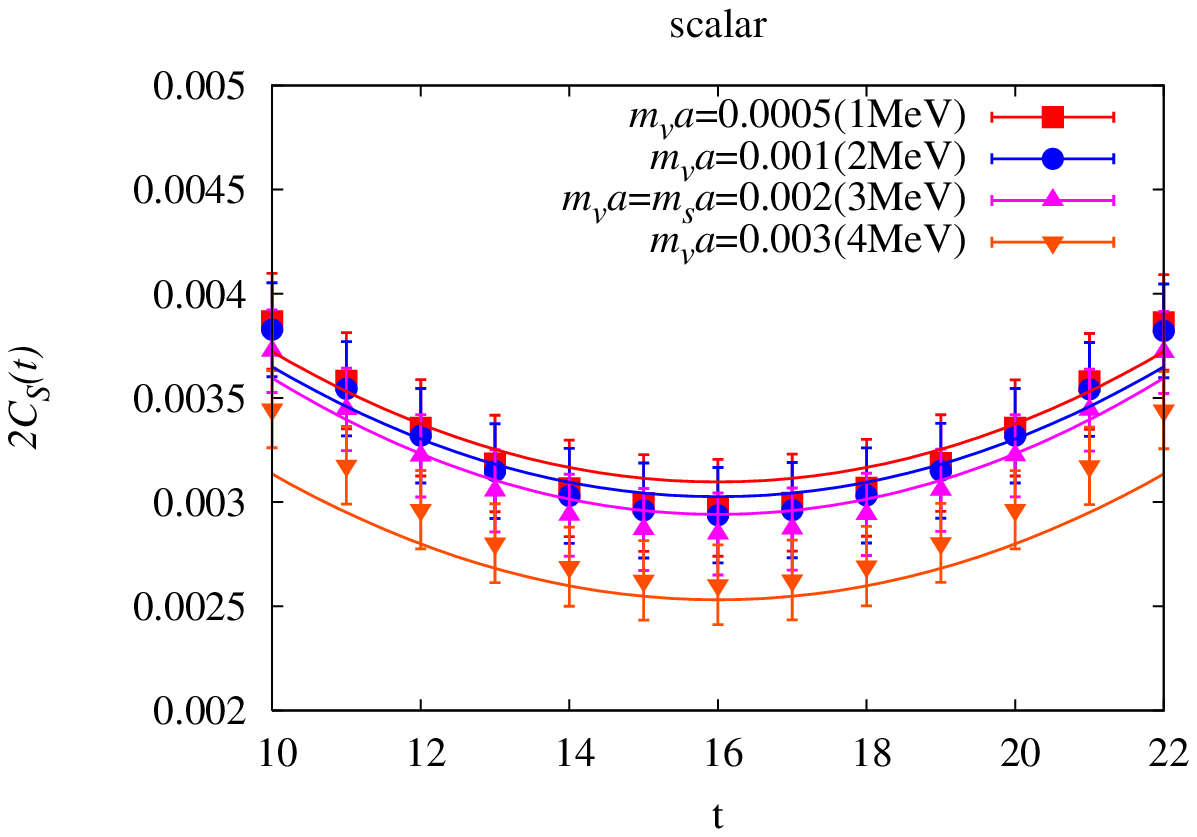}
  \caption{
    Partially quenched pseudo-scalar (top) and scalar (bottom) 
    correlators in the $\epsilon$-regime.
    The solid curves are the ChPT results with 
    $\Sigma = [227.6(3.7)\mbox{MeV}]^3$ and
    $F = 87.3(5.6)\mbox{MeV}$ as inputs (No free parameter left).
  }
  \label{fig:PQChPT1}
\end{figure}

\begin{figure}[tbp]
  \centering
  \includegraphics[width=12cm]{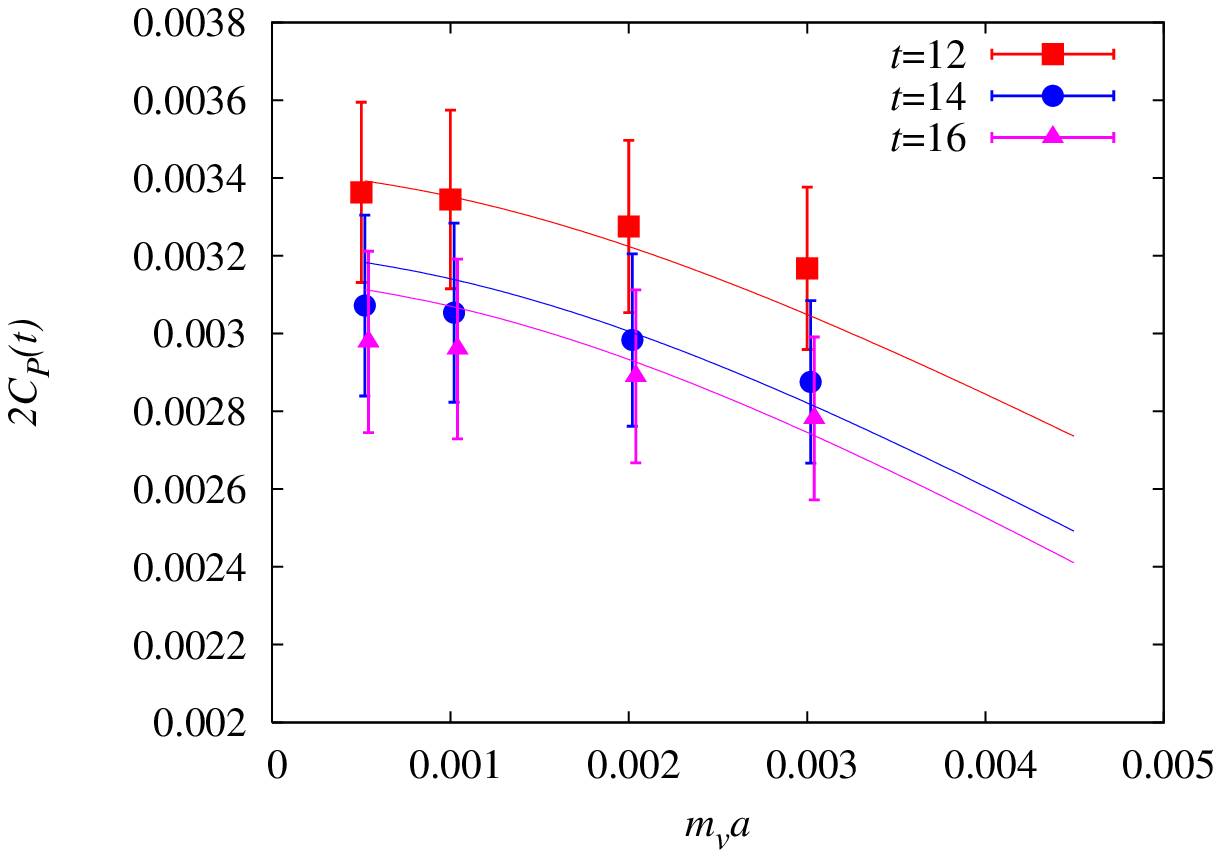}
 \includegraphics[width=12cm]{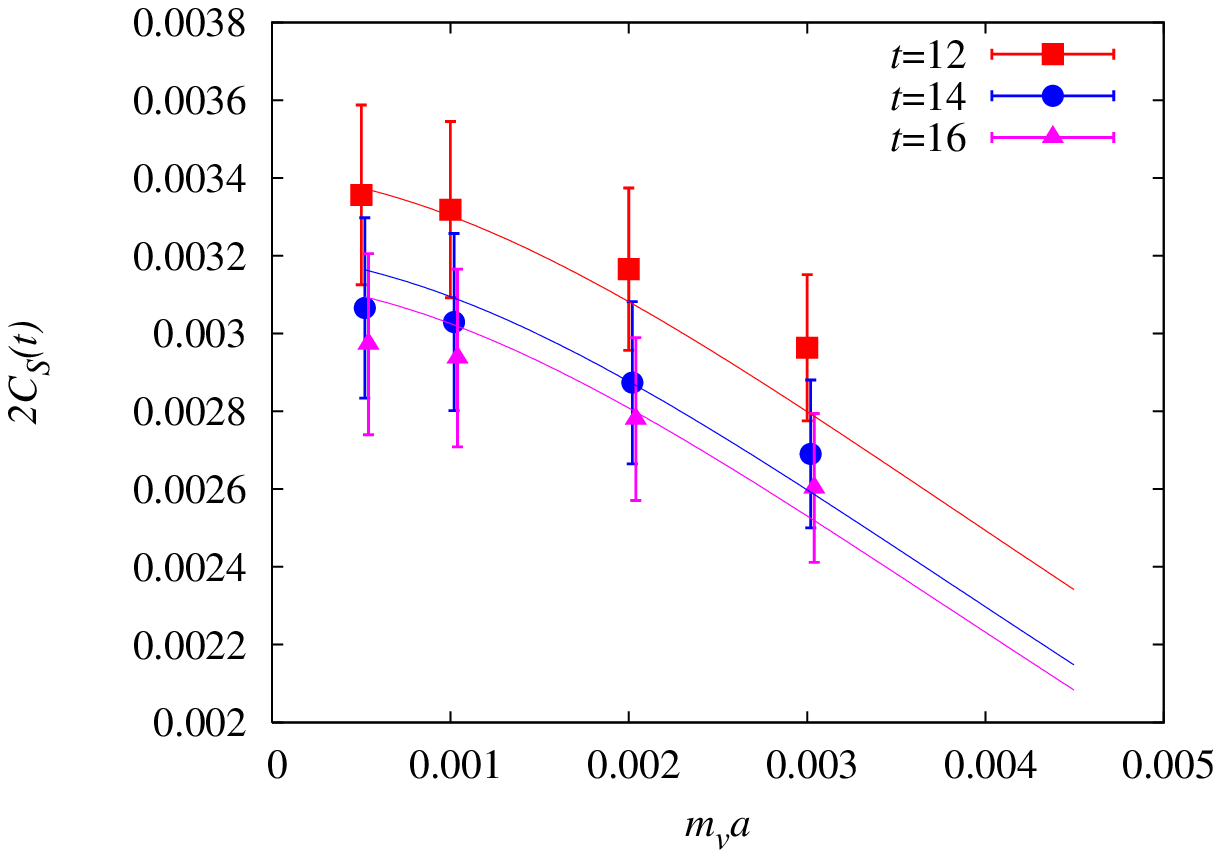}
  \caption{
    Valence quark mass dependence of the partially quenched pseudo-scalar
    (top) and scalar (bottom) correlators in the $\epsilon$-regime at each
    time slice. 
  }
  \label{fig:PQChPT2}
\end{figure}

Using the data at $m_v\not=m_s$ we can check the consistency of the lattice
results with the partially quenched ChPT formulae (\ref{eq:Pmult}) and
(\ref{eq:Smult}). 
Figure~\ref{fig:PQChPT1} shows the pseudo-scalar and scalar correlators at
four different valence quark masses ranging 1--4~MeV while fixing the sea
quark mass at 3~MeV.
The shape of the correlators does not strongly depend on the valence quark
mass, but some dependence can be seen.
With the input parameters $F$ and $\Sigma$ determined from a global fit of the
axial-vector and pseudo-scalar channel (see Section~\ref{sec:LEC}),
we draw the expectation from the partially quenched ChPT in
Figure~\ref{fig:PQChPT1}.
There is no additional parameters in this
analysis, hence it gives a stringent test of either the NLO ChPT formula or
the lattice calculation.
We find an excellent agreement for all the valence quark masses within the
statistical error.
Some specific points of the correlator, {\it e.g.} at $t$ = 12, 14, or 16, are
plotted as a function of the valence quark mass in Figure~\ref{fig:PQChPT2}.
The decrease of the correlators with the valence quark mass is nicely
reproduced by the lattice data.

\section{Extraction of the low energy constants}
\label{sec:LEC}

Since the axial-vector and pseudo-scalar channels are complementary to each
other in determining the LECs $F$ and $\Sigma$, we fit both channels
simultaneously with these two parameters to obtain our best results:
$\Sigma=[227.6(3.7)\mathrm{~MeV}]^3$ and
$F=87.3(5.6)\mathrm{~MeV}$.
Here, the statistical error of the lattice spacing $a$
is also taken into account.
The fit range is again $t\in [12,20]$ for both channels.
Under the change of the lower limit of the fitting range from 10 to 15, 
which corresponds to 1.1~fm to 1.7~fm, 
the fit results are quite stable (within 1\%) with similar error bars.

Multiplying the non-perturbative renormalization factor
\cite{Martinelli:1994ty}
to convert our result of the chiral condensate
to the continuum $\overline{\mathrm{MS}}$ scheme
we obtain 
$\Sigma^{\overline{\mathrm{MS}}}(\mathrm{2~GeV}) = [239.8(4.0)\mathrm{~MeV}]^3$,
where the error represents the statistical one.

Possible sources of the systematic error include the discretization effect of
$O(a^2)$ and the higher order effect of the $\epsilon$-expansion.
Since we do not have data at different lattice spacings, we cannot quantify
the discretization effect beyond a rough order counting.
Assuming that the relevant physical scale is the QCD scale
$\Lambda_{\mathrm{QCD}}\sim$ 300--500~MeV, the size of
$\mathcal{O}((a\Lambda_{\mathrm{QCD}})^2)$ effect is 3--8\%.
This small scaling violation is supported for the overlap fermions by recent
quenched simulations (in the $p$-regime) \cite{Babich:2006bh,Draper:2006wb},
albeit for different physical quantities.
The unphysical (heavy) Wilson fermions we introduced to fix the topological
charge should not have any negative impact on the scaling, because they never
arise in the external states and merely affect the gluon action at order $a^2$
in the Symanzik's effective theory.
Their effect on the effective gluon action is further minimized by the
accompanied ghosts that cancel the Wilson fermion contributions except for the
near-zero modes of $H_W$, that appear for locally bumpy gauge configurations,
or for the so-called dislocations \cite{Golterman:2003qe}. 

The higher order effect of the $\epsilon$-expansion appears due to the finite
volume lattice.
In our analysis, the NLO terms, {\it i.e.} $\mathcal{O}(\epsilon^2)$ terms, are
included and the remaining corrections are of
$\mathcal{O}(\epsilon^4)\sim\mathcal{O}(1/(\Lambda_{\mathrm{cut}}L)^4)$,
with $\Lambda_{\mathrm{cut}}$ the cutoff scale of ChPT.
With a conservative choice $\Lambda_{\mathrm{cut}}\sim$ 300--500~MeV, this
uncertainty is 0.3--2\%.
This order-counting gives a reasonable estimate for the NLO correction to
$\Sigma$. 
Namely, the size of the NLO correction $(\Sigma_{\rm eff}-\Sigma)/\Sigma$
(see (\ref{eq:Sigmaeff})) estimated with this order-counting is 14\%, 
while the real correction was 20\%.
This agreement indicates that the order-counting argument is indeed valid
up to an $O(1)$ factor, which depends on the shape of the space-time box  
\cite{Hasenfratz:1989pk} and could be sizable at the NNLO.
At the same order of the $\epsilon$-expansion, there is a contribution from
the finite pion mass $\mathcal{O}((m_\pi/\Lambda_{\mathrm{cut}})^2)$, which
numerically gives about 2\%.
We note that such a small uncertainty does not apply for the curvature of the
correlators, since the curvature itself is the quantity arising at the NLO. 
But the extraction of $F$ and $\Sigma$ relies mainly on the constant piece of
the correlators, that starts from the leading order in the
$\epsilon$-expansion.

Adding the uncertainties from the scaling violation and from the
higher order effects in the $\epsilon$-expansion 
in quadrature 
we estimate 
the dominant systematic error is of order 9\%. 
For more robust estimate beyond the order-counting, it is important to
study the finite lattice spacing and volume effects in the future works.

The result for the decay constant $F$ may be compared with the calculation in
the $p$-regime \cite{Necco:2007pr}.
In \cite{Noaki:2007es} a preliminary result for the pion decay constant with
two-flavors of dynamical overlap fermion is presented.
With the same lattice action, the calculation is done at slightly coarser
lattice spacing, $a\simeq$ 0.12~fm.
An analysis with the next-to-next-to-leading order (NNLO) ChPT yields
$F=78.6(2.7)$~MeV (the error denotes the statistical one), 
which is slightly lower than the calculation in this work.
Assuming the 9\% systematic error, however, both results are consistent with
each other.

The result for the chiral condensate $\Sigma$ may be compared with the
extraction from the lowest-lying eigenvalue through the chiral Random Matrix
Theory \cite{Fukaya:2007fb,Fukaya:2007yv}:
$\Sigma^{\overline{\mathrm{MS}}}(2\mathrm{~ GeV}) = [251(7)(11)\mathrm{~MeV}]^3$.
Here, the errors represent the statistical and the systematic due to the NLO
correction in the $\epsilon$-expansion.
The calculation in this work from the meson correlators includes the NLO
correction, and in fact the result deviates from the previous result by about
the estimated NLO correction.
When one integrates out the non-zero modes of the pion field from the chiral
Lagrangian, there remains the zero-mode integrals with an ``effective'' 
chiral condensate, which is given as
\begin{equation}
\Sigma_{\rm{eff}}=
\Sigma \left(1 + \frac{3}{2}\;\frac{0.0836}{F^2L^2} \right)
\end{equation}
at the one-loop order.
The one-loop correction is a substantial effect\cite{Damgaard:2007xg}: 
a factor of 1.202.
If we correct the value in this work with this amount we obtain 
$[255(4)\mathrm{~MeV}]^3$, which is now consistent with the previous result
within the small statistical error.
This remarkable consistency has already been discussed in 
\cite{DeGrand:2006nv,Hasenfratz:2007yj,DeGrand:2007tm,DeGrand:2007mi},
and an argument based on an analytical calculation has been given recently
\cite{Basile:2007ki}.
It is also notable that our results are consistent with
the topological susceptibility, from which we extracted
$\Sigma^{\overline{\mathrm{MS}}}(2\mathrm{~ GeV}) = [254(5)(10)\mathrm{~MeV}]^3$
(See \cite{Aoki:2007ka, Aoki:2007pw} for the details). 

\section{Conclusion}
\label{sec:conclusion}

The extraction of the low-energy constants $F$ and $\Sigma$ from the meson
correlators in the $\epsilon$-regime has a significant advantage over the
conventional approach.
Already at the NLO in the $\epsilon$-expansion, the remaining higher order
effect is a per cent level, and thus a precise calculation of both $F$ and
$\Sigma$ is possible without delicate chiral extrapolations.

The dynamical lattice simulation is feasible with a small sea quark mass
$\sim$ 3~MeV on a $16^3\times 32$ lattice with $L\sim 1.7$~fm,
where the scale is determined assuming $r_0 = 0.49$ fm.
Meson correlators are obtained with a good precision by using the low-mode
averaging technique.
The numerical results passed all the consistency checks for different channels
as well as for the partially quenched mass combinations.
Using the NLO ChPT formulae in the $\epsilon$-expansion, we obtain
$\Sigma^{\overline{\mathrm{MS}}}(\mathrm{2~GeV}) = [239.8(4.0)\mbox{~MeV}]^3$
and $F = 87.3(5.6)$~MeV
with a possible systematic error of order 9\% which is 
dominated by the discretization error.
Reducing the systematic error by repeating the calculation
on a larger lattice at smaller lattice spacing will be an
important future work.

\section*{Acknowledgments}
HF thanks F.~Bernardoni, P.~H.~Damgaard, T.~DeGrand, L.~Giusti, 
A.~Hasenfratz, P.~Hasenfratz, P.~Hern\'andez, S.~Necco
and K.~Splittorff
for fruitful discussions.
The numerical simulations are performed 
on IBM System Blue Gene Solution at High 
Energy Accelerator Research Organization (KEK)
under support of its Large Scale Simulation
Program (No.07-16), also in part on NEC SX-8 at YITP, Kyoto University.
This work is supported in part by Nishina foundation (HF) and by
the Grants-in-Aid for
Scientific Research from the Ministry of Education,
Culture, Sports, Science and Technology.
(Nos. 17740171, 18034011, 18340075, 18740167, 18840045,19540286 and 19740160). 


%

\end{document}